\documentclass[aps,prd,twocolumn,groupedaddress,nofootinbib]{revtex4}

\usepackage{amsfonts,amsmath,amssymb,mathrsfs}
\usepackage{hyperref}
\usepackage{color}
\usepackage{graphicx}  
\usepackage{dcolumn}   
\usepackage{bm}        
\usepackage[english]{babel}
\def\a{\alpha}

\def\r{\rho}
\def\s{\sigma}
\def\t{\tau}
\def\m{\mu}
\def\n{\nu}
\def\k{\kappa}
\def\th{\theta}
\def\g{\gamma}\def\G{\Gamma}
\def\L{\Lambda}\def\l{V}
\def\D{\Delta}
\def\la{\langle}
\def\ra{\rangle}
\def\o{\omega}\def\O{\Omega}
\def\d{\delta}
\def\p{\partial}

\def\oxthree{{\cal O}(x^3) }

\def\half{\textstyle{\frac{1}{2}}}

\def\bdoc{\begin{document}}
\def\edoc{\end{document}}
\def\bea{\begin{equation}}
\def\eea{\end{equation}}

\def\beq{\begin{eqnarray}}
\def\eeq{\end{eqnarray}}
\def\ben{\begin{enumerate}}
\def\een{\end{enumerate}}
\def\la{\langle}
\def\ra{\rangle}
\def\a{\alpha}
\def\g{\gamma}\def\G{\Gamma}
\def\d{\delta}\def\D{\Delta}
\def\e{\epsilon}
\def\z{\zeta}

\def\th{\theta}
\def\k{\kappa}
\def\l{\lambda}
\def\m{\mu}
\def\n{\nu}
\def\o{\omega}
\def\p{\pi}
\def\r{\rho}
\def\s{\sigma}
\def\t{\tau}
\def\L{{\cal L}}
\def\S{\Sigma }
\def\gsim{\; \raisebox{-.8ex}{$\stackrel{\textstyle >}{\sim}$}\;}
\def\lsim{\; \raisebox{-.8ex}{$\stackrel{\textstyle <}{\sim}$}\;}
\def\gtrsim{\gsim}
\def\lessim{\lsim}
\def\loc{{\rm local}}
\def\vm{v_{\rm max}}
\def\bh{\bar{h}}
\def\del{\partial}
\def\nab{\nabla}
\def\half{{\textstyle{\frac{1}{2}}}}
\def\fourth{{\textstyle{\frac{1}{4}}}}

\def\bD{{\bf D}}
\def\bE{{\bf E}}
\def\bF{{\bf F}}
\def\bB{{\bf B}}
\def\bP{{\bf P}}
\def\bV{{\bf v}}
\def\bv{{\bf v}}
\def\bx{{\bf x}}
\def\by{{\bf y}}
\def\bz{{\bf z}}
\def\ba{{\bf a}}
\def\bd{{\bf d}}
\def\bs{{\bf s}}
\def\bn{{\bf n}}
\def\bp{{\bf p}}

\def\O{\Omega}

\def\br{{\bf r}}
\def\bnab{{\bf \nab}}

\def\tE{\tilde{E}}
\def\tL{\tilde{L}}
\def\Horava{Ho\v{r}ava }

\def\oxtwo{\mathscr{O}\left(x^2\right)}
\def\oxthree{\mathscr{O}\left(x^3\right)}
\def\oxfour{\mathscr{O}\left(x^4\right)}
\def\oxfive{\mathscr{O}\left(x^5\right)}

\def\ph{\phantom}
\def\LL{Lanczos-Lovelock}

\begin{document}
\title{Thermodynamics of Local Causal Horizons}
\author{Arif Mohd}
 \email{arif.mohd@sissa.it}
\affiliation{SISSA - International School for Advanced Studies, Via Bonomea 265, 34136 Trieste, Italy \\
and \\
INFN, Sezione di Trieste}
\author{Sudipta Sarkar}
\email{sudiptas@iitgn.ac.in}
\affiliation{Indian Institute of Technology, Gandhinagar, India}
\begin{abstract}
We propose an expression for the entropy density associated with the Local Causal Horizons in any diffeomorphism invariant theory of gravity. If the black-hole entropy of the theory satisfies the physical process version of the first law of thermodynamics then our proposed entropy satisfies the Clausius relation. Thus, our study shows that the thermodynamic nature of the spacetime horizons is not restricted to the black holes; it also applies to the local causal horizons in the neighborhood of any point in the spacetime.
\end{abstract}
\maketitle
Black-hole physics has provided strong hints of a deep and fundamental relationship between gravitation, thermodynamics and quantum theory.
At the heart of this relationship is black-hole thermodynamics, which says that certain laws of black hole mechanics are, in fact, simply the
ordinary laws of thermodynamics applied to a system containing a black
hole \cite{Bardeen:1973gs}. Classical and semi-classical analyses of the thermodynamic behavior of black holes has given rise to
most of our present physical insights into the nature of quantum phenomena
occurring in strong gravitational fields. 
\par
The equilibrium state version of the first law of black-hole thermodynamics for arbitrary diffeomorphism-invariant theory of gravity was established by Wald and collaborators \cite{Wald:1993nt,Iyer:1994ys}. 
It is also possible to write down a quasi-stationary version of the second law for \LL\ gravity
\cite{Chatterjee:2011wj, Kolekar:2012tq} for physical processes in which the horizon is perturbed by the
accretion of positive energy matter and the black hole ultimately settles down to a stationary state.\par
The discovery of the black-hole thermodynamics was also a precursor to the idea of emergent gravity, which asserts that spacetime is an emergent notion, a macroscopic and coarse-grained description of the underlying degrees of freedom which are fundamentally described by a quantum theory of gravity. But how could gravitation be an emergent phenomenon if the thermodynamics is tied to special spacetimes - those containing a black hole? In this direction a profound insight was provided by Jacobson who managed to derive the Einstein equation as a consequence of the thermodynamics associated to the Local Causal Horizons (LCHs) \cite{Jacobson:1995ab}. Jacobson showed that if one identifies a multiple of area with the entropy $(S \propto A)$ associated with a LCH at boost temperature $T$, and if the change in the entropy $\delta S$ is related to the flow of the boost energy $\delta Q$ across the local horizon in such a way that the Clausius relation $\delta S=\delta Q / T$ is satisfied, then it is possible to derive the Einstein equation. The Einstein equation can therefore be interpreted as a thermodynamic equation of state. \par
This simple and elegant result has far reaching consequences. It supports the idea that the dynamics of space time can be interpreted as a thermodynamic description of some underlying degrees of freedom.
Also, as pointed out by Padmanabhan \cite{Padmanabhan:2009vy}, one can deduce the presence of these underlying degrees of freedom by the very observation that one can heat the system. The existence of physical quantities which provide a thermodynamical interpretation of the geometry of spacetime and the existence of a thermodynamical law  (the Clausius relation) that gives rise to the dynamics of the geometry (the Einstein equation) is highly suggestive of the idea that the dynamical geometry itself could be an emergent notion borne out of the dynamics of some underlying degrees of freedom.  \par
Now, the Einstein-Hilbert action has just the first of the terms in a derivative expansion and one expects {\it \`a la} Wilson that higher-order curvature terms respecting the symmetries should also appear in the effective action. It is natural to ask if the thermodynamical nature of LCHs is unique to general relativity or is it a robust feature of any diffeomorphism invariant theory of gravity. Although there have been many attempts \cite{Eling:2006aw, Chirco:2010sw, Elizalde:2008pv, Brustein:2009hy, Parikh:2009qs, Guedens:2011dy} to answer this question and extend the result of Ref.~\cite{Jacobson:1995ab} to other theories of gravity, none of these are completely satisfactory. Either they work only for a very specific class of theories or they require extra assumptions. A detailed critique and  the limitations of these attempts is discussed in Ref.~\cite{Guedens:2011dy}. \par 
In this paper we take the point of view that if the thermodynamics of LCHs is a fundamental property of a theory of gravity and there are internal degrees of freedom whose coarse-grained description is provided by the low energy gravitational effective action then we must be able to show the existence of thermodynamical laws associated to the LCHs with appropriately defined quantities as the thermodynamic variables.  
Our goal is opposite to the one that Jacobson had in Ref.~\cite{Jacobson:1995ab}. Instead of deriving the field equation of gravity from the assumed thermodynamics associated to the LCHs, we use the field equations to derive the Clausius relation for the LCHs. More precisely, for any given diffeomorphism invariant theory of gravity in which black holes satisfy the physical process version of the first law of thermodynamics, we propose an expression of the entropy associated with the LCHs that satisfies the Clausius relation. \par
This paper is organized as follows: in Sec.~\ref{Geometric setup} we discuss the geometrical construction of our LCH. In this section we also review the derivation of the Clausius relation in general relativity and the problem one faces in general theories of gravity. In Sec.~\ref{our proposal} we present our proposal of the entropy density associated with the LCHs in arbitrary theories of gravity. Summary and outlook are presented in Sec.~\ref{discussion}.

\section{Geometric setup}
\label{Geometric setup}
\begin{figure}
\includegraphics[scale=0.35]{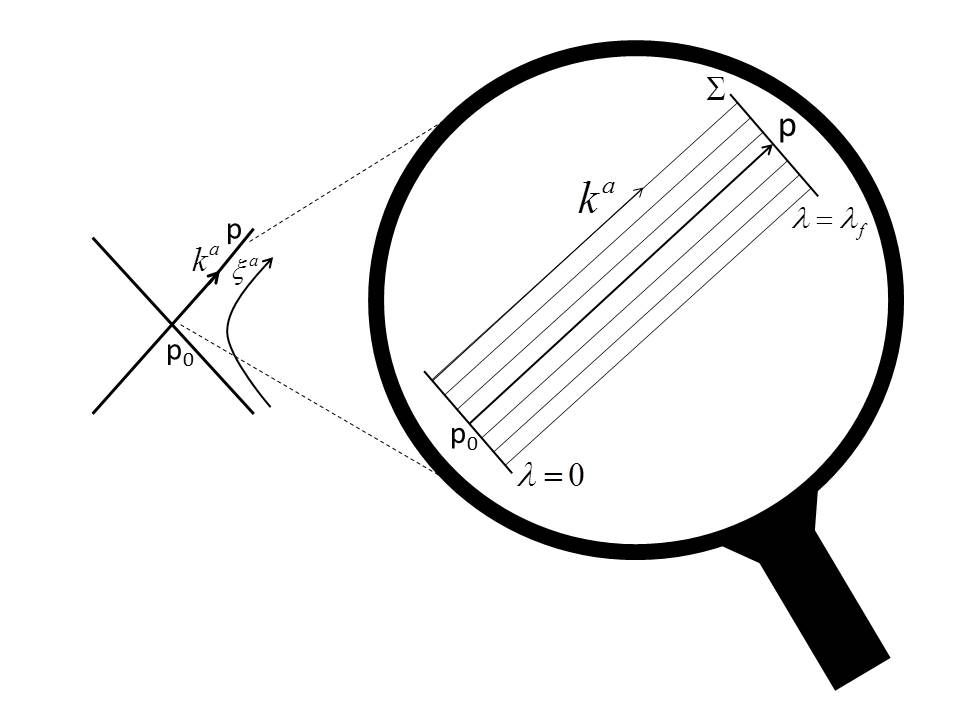}
\caption{Local Causal Horizon (LCH) associated with the approximate boost Killing vector field $\xi^a$. $p_0$ is the bifurcation point where $\xi^a$ vanishes. $k^a$ is the affinely parametrized tangent to the null generators of the LCH with the affine parameter $\lambda$. $p$ is the final equilibrium point where the expansion $\theta$ and shear $\sigma$ of $k^a$ vanishes. The affine parameter vanishes at $p_0$ and takes a value $\lambda_f$ at $p$. On the generator connecting $p_0$ to $p$ we have $\xi^a = \lambda k^a$.}\label{fig:lch}
\end{figure}

 We start with the construction of the LCH based at any point $p$ in spacetime. We first choose a  $(D-2)$-dim spacelike surface ${\Sigma}$ containing $p$ and we choose one side of the boundary of the past of ${\Sigma}$. In a small neighborhood of $p$ this boundary is generated by a congruence of null geodesics $k^a$ normal to ${\Sigma}$. Let $\lambda$ be the affine parameter along the integral curves of $k^a$ such that $\lambda = \lambda_f$ at $p$. Our choice of the patch ${\Sigma}$ is such that the expansion $\theta$ and the shear $\sigma_{ab}$ of $k^a$ vanish at $p$. Our chosen side of the null boundary of the past of such a ${\Sigma}$ is called the local causal horizon (LCH)\footnote{Our construction of the LCH is different from the one originally given by Jacobson \cite{Jacobson:1995ab}. There the equilibrium was at the bifurcation point (where the Killing vector vanishes), while for us the equilibrium point ($p$) lies in the future of the bifurcation point ($p_0$). Both of these constructions are discussed in Ref.~\cite{Guedens:2011dy}.} at $p$, see Fig.~\ref{fig:lch}. \par
In order to define the heat flux and the temperature we need an approximate boost Killing vector field. Consider a point $p_0$ in the past of $p$ lying at the value of the affine parameter $\lambda = 0$ on the central generator (i.e., the generator that passes through $p$). Next we construct an approximate boost Killing vector field $\xi^a$ such that the point $p_0$ serves as a bifurcation point, i.e., $\xi^a$ vanishes at $p_0$.  On the central generator the approximate boost Killing vector is related to the null-geodesic generator as $\xi^a = \lambda k^a$. Furthermore, $\xi^a$ satisfies the Killing equation $\nabla_{(a} \xi_{b)} = 0$ exactly at $p$, and to $\mathcal{O}(x)$ near $p$. We suggest the interested reader to consult the references \cite{Guedens:2011dy,Guedens:2012sz} for further details pertinent to this construction. The appropriate temperature is $T = 1 / 2\pi$ which is the boost temperature associated with the approximate Killing vector $\xi^a$. The heat flux is
given in terms of the matter stress tensor and the approximate boost Killing vector field $\xi^a$ as
\begin{eqnarray}
\label{eq:heat flux}
\delta Q &=& \int_{0}^{\lambda_f} \, d\lambda \, \,dA \, \sqrt{\gamma}\,\, T_{ab} \xi^a k^b \nonumber \\
&=& \int_{0}^{\lambda_f} \,\lambda \,\, d\lambda \, \,dA \, \sqrt{\gamma}\,\, T_{ab} k^a k^b.
\end{eqnarray}
Here $\gamma$ is the induced metric on the LCH and the $dA$ integral is over a thin pencil of generators around the central generator connecting $p$ to $p_0$. \par
We also need an entropy associated with the LCH. To begin with, taking a clue from the black-hole thermodynamics, it seems reasonable to guess that one should associate the same expression of entropy to LCHs as that of the black holes in the theory. Let us then assume that the underlying theory of gravity obeys an equation $E_{ab} = 8 \, \pi\, T_{ab}$ and the entropy of any slice of LCH is same as that of the black-hole horizon in that theory given by,
\begin{eqnarray}
 S = \frac{1}{4} \int \,  dA \, \sqrt{\gamma}\,\rho_{B},
\end{eqnarray}
where $\rho_{B}/4$ is the entropy density of the black-hole horizon. For example, in general relativity $\rho_{B}$ would be equal to $1$. Let us now consider how this entropy changes along the null congruence $k^a$. The entropy change is
\beq
\delta S &=& \frac{1}{4} \int_{0}^{\lambda_f} \, d\lambda \, dA \, \sqrt{\gamma} \left( \rho_{B} \, \theta
+ \frac{d \rho_{B}}{d \lambda} \right).
\eeq
Next, we expand the entropy change in a Taylor series around the terminal cross section at $\lambda = \lambda_f$, 
\begin{multline}
\label{eq:delta S old}
\delta S = \frac{1}{4} \int_{0}^{\lambda_f} \, d\lambda \, dA \, \sqrt{\gamma} \biggl[  \left( \rho_{B} \, \theta + \frac{d \rho_{B}}{d \lambda} \right)_{\lambda_f}  +  \\
(\lambda -\lambda_f ) \frac{d}{d\lambda}\left( \rho_{B} \, \theta + \frac{d\rho_{B}}{d \lambda} \right)_{\lambda_f} + \,\,\, \dots \biggr],
\end{multline}
 where the suffix $\lambda_f$ indicates that the quantity in the bracket is evaluated at the final cross section $\lambda = \lambda_f$. Since $\int_{0}^{\lambda_f} \, d\lambda \, \lambda$ appearing in $\delta Q$ in Eq.~\eqref{eq:heat flux} is of the same order in $\lambda_f$ as $\int_{0}^{\lambda_f} \, d\lambda \, (\lambda - \lambda_f)$, the second term of expansion in Eq.~\eqref{eq:delta S old}, for the validity of the Clausius relation we must have that the first term of expansion in Eq.~\eqref{eq:delta S old} vanishes. By our construction of LCH, $\theta$ is zero at $\lambda_f$ on the central generator.  Hence a necessary condition for the validity of the Clausius relation for the LCH is that
\beq 
\left(\frac{d \rho_{B}}{d\lambda}\right)_{\lambda_f} = \, 0.\label{condition_1}
\eeq
In general relativity $\rho_{B}$ is $1$ and Eq.~\eqref{condition_1} is trivially satisfied. Hence, for general relativity we get from 
Eq.~\eqref{eq:delta S old},
\begin{align}
\label{eq:delta S in gr}
\delta S 
 =  \frac{1}{4} \int_{0}^{\lambda_f} \, d\lambda \, dA \, \sqrt{\gamma} (\lambda -\lambda_f ) \left(- R_{ab} k^a k^b  \right)_{\lambda_f},
\end{align}
where 
we have used the Raychaudhari equation and the fact that $\theta(\lambda_f) = 0 = \sigma(\lambda_f)$. Now using the Einstein equation and doing the $\lambda$ integral we see that Eq.~\eqref{eq:delta S in gr} is equal to $\delta Q/T$, where $\delta Q$ is as in Eq.~\eqref{eq:heat flux} and $T$ is the local Unruh Temperature, $T = 1/2\pi$. Hence it is easy to prove the Clausius relation for LCHs in general relativity. For any other theory of gravity, however, the situation is different. For example, consider a  theory of gravity described by the Lagrangian which is a function of the Ricci scalar $f(R)$. The Wald entropy density associated with the black holes of such a theory is $ \sim f'(R)$, and the validity of Eq.~\eqref{condition_1} requires that
$
\left({d R}/{d\lambda}\right)_{\lambda_f} = \, 0.  
$
But the spacetime is completely arbitrary and in general this is not true on the final cross section of the LCH. In comparison, in the physical-process version of the first law \cite{Jacobson:2003wv, Amsel:2007mh, Chatterjee:2011wj} for black holes,  the final cross-section of the horizon is part of a stationary spacetime, and therefore, all changes of the dynamical fields vanish in the asymptotic future. In particular, we can set $ dR / d\lambda = 0$ on the final cross-section. \par
This is the crucial geometric difference between the case of black holes and that of the LCHs. The future boundary condition for black holes is physically motivated from the stability and cosmic censorship hypothesis \cite{Wald:1984rg}. There is no compelling reason to impose such a condition on the LCHs. This is the difficulty in imposing Eq.~\eqref{condition_1} necessary for deriving the Clausius relation for the LCHs. \par 
It was suggested in  Ref.~\cite{Eling:2006aw} that instead of requiring that $\theta  = 0$ at $p$ one could
choose the patch $\Sigma$ such that the first term of the expansion
in Eq.~\eqref{eq:delta S old}  vanishes identically at $p$. Then one could derive a non-equilibrium version of the Clausius relation that contains an extra term which is
interpreted as the entropy-production term. In Ref.~\cite{Chirco:2010sw},
this entropy production term is shown to be a separate
contribution to the heat 
flux coming from the additional
scalar degree of freedom present in f(R) gravity. However, it is explained in Ref.~\cite{Guedens:2011dy} why such an approach may not work beyond the simple case
of f(R) gravity. 
\section{A new proposal for the entropy density}
\label{our proposal}
In this paper we follow a different approach and 
we seek a new candidate for the entropy density associated with the LCHs. There is some minimal set of properties one expects from this entropy density: it should of course satisfy the Clausius relation; as in the black-hole case it is desirable to have $\theta=0$ and $\sigma_{ab} =0$ as the definition of the equilibrium slice of the LCH; for a stationary black hole the proposed entropy should agree with the entropy which satisfies the physical-process version of the first law for black holes. 
With these guidelines in mind, our proposal is that the entropy density $\rho$ of any slice $\Sigma_\lambda$ of LCH, located between $p$ and $p_0$ at an affine parameter $\lambda \in [0,\lambda_f]$, is given by
\begin{multline}
\label{eq:rho}
\rho = \rho_{B} + (\lambda_f - \lambda) \left(\frac{d \rho_{B}}{d \lambda}\right)_{\lambda_f} + \\
\frac{(\lambda_f - \lambda)^2}{2}  \left(\rho_{B} R_{ab} k^a k^b - \frac{d^2 \rho_{B}}{d \lambda^2} - E_{ab} k^a k^b\right)_{\lambda_f}, 
\end{multline}
where $\rho_{B} /4$ represents the expression for the entropy density of the stationary black holes of the theory in question. The entropy of the slice is given by
\begin{align}
S = \frac{1}{4} \int_{\Sigma_\lambda} dA \sqrt{\gamma} \, \rho.
\end{align}
Note that the entropy density of the final equilibrium slice is equal to $\rho_{B}$, the entropy of the stationary black hole in the theory. \par
Let us now check if our expression of the entropy density leads to the correct thermodynamics of the LCH. The change in entropy as we evolve the system from the initial slice at $\lambda = 0$ to the final equilibrium slice at $\lambda = \lambda_f$ is 
given by Eq.~\eqref{eq:delta S old}, with $\rho_{B}$ now replaced by $\rho$ in Eq.~\eqref{eq:rho}, as
\begin{multline}
\label{eq:delta S new}
\delta S = \frac{1}{4} \int_{0}^{\lambda_f} \, d\lambda \, dA \, \sqrt{\gamma} \biggl[  \left( \rho \, \theta + \frac{d \rho}{d \lambda} \right)_{\lambda_f}  +  \\
(\lambda -\lambda_f ) \frac{d}{d\lambda}\left( \rho \, \theta + \frac{d\rho}{d \lambda} \right)_{\lambda_f} + \,\,\, \dots \biggr].
\end{multline}
Now, noticing that $\theta(\lambda_f)=0$ because of our equilibrium condition, the first term in the big square brackets just gives $\left({d \rho}/{d\lambda}\right)_{\lambda_f}$ which is easily seen to be equal to $0$ from Eq.~\eqref{eq:rho}. The coefficient of $(\lambda - \lambda_f)$ can be calculated in a straightforward fashion using the Raychaudhari equation and it gives simply $(-E_{ab} k^a k^b )_{\lambda_f}$. In the limit that the initial slice gets very close to the final equilibrium slice we get,
\begin{align}
\label{eq:delta S new}
\delta S &= \frac{1}{4}\int_{\Sigma_{\lambda_f}} dA \sqrt{\gamma} E_{ab}k^a k^b \int_{0}^{\lambda_f} d\lambda (\lambda_f- \lambda)  \nonumber \\
& = \frac{\lambda_f^2}{2}\,\, \frac{1}{4}\int_{\Sigma_{\lambda_f}} dA \sqrt{\gamma} E_{ab}k^a k^b.
\end{align}
In the same limit we get, from Eq.~\eqref{eq:heat flux} for the total heat flux through the null surface enclosed between $\lambda = 0$ and $\lambda= \lambda_f$,
\begin{align}
\label{eq:delta Q new}
\delta Q = \frac{\lambda_f^2}{2} \, \int_{\Sigma_{\lambda_f}} \,dA \, \sqrt{\gamma}\,\, T_{ab} k^a k^b.
\end{align}
Now using the local boost temperature $T = 1/2\pi$ and the equation of motion $E_{ab} = 8 \pi T_{ab}$, we see from equations~\eqref{eq:delta S new} and \eqref{eq:delta Q new} that the Clausius relation $\delta S=\delta Q / T$ is satisfied at the LCH. \par
To appreciate the non-trivial nature of our entropy density, it is instructive to see an explicit expression of $\rho$ for some theories. In $f(R)$ gravity the black-hole entropy is known to be given by $f'(R)$ \cite{Elizalde:2008pv}. Using the equation of motion for $f(R)$ gravity,
\begin{align}
\label{eom f(R)}
f'(R) R_{ab} - \nabla_a \nabla_b f'(R) + (\Box f'(R)  &- \frac{1}{2}f'(R)) g_{ab} \\ \nonumber
&= 8 \pi G \,T_{ab},
\end{align}
and the black-hole entropy, $f'(R)$, we see that the coefficient of $(\lambda-\lambda_f)^2$ in Eq.~\eqref{eq:rho} vanishes and we are left with the following expression for the entropy density of LCHs in $f(R)$ gravity,
\begin{align}
\label{eq:rho f(R)}
\rho=f'(R)+(\lambda_f - \lambda) \left(\frac{d f'(R)}{d \lambda}\right)_{\lambda_f}.
\end{align}
We emphasize that with our construction of the LCH and the definition of $\rho$ there is no need for the non-equilibrium entropy-production terms in the Clausius relation as in Ref.~\cite{Eling:2006aw}. As shown above, this entropy density satisfies the Clausius relation without any extra terms. \par
Finally, we also give here $\rho$ for the Einstein-Gauss-Bonnet (EGB) gravity although the expression is not particularly illuminating.  Action of the EGB theory in D-dimensions is given by 
\begin{align}
\label{EGB action}
S_{EGB} = \frac{1}{16 \pi G} \int d^D x \sqrt{-g}\,(R+\alpha \mathcal{L}_{GB}),
\end{align}
where $\mathcal{L}_{GB}=R^2 - 4 R_{ab}R^{ab}+R_{abcd}R^{abcd}$. The field equation is $G_{ab} + \alpha H_{ab}=8\pi G \,T_{ab}$, where $G_{ab}$ is the Einstein tensor and $H_{ab}$ is given by
\begin{align}
H_{ab} = 2 ( R R_{ab} - 2 R_{am} R^{mb} &- 2 R^{pq}R_{apbq}+\nonumber \\
&R_a^{pqr}R_{bpqr}) -\frac{1}{2}g_{ab}\mathcal{L_{GB}}. \nonumber
\end{align}
The entropy of black holes in the EGB thoery is given by $\rho_{B} = 1+2\alpha \,\hat{R}$, where $\hat{R}$ is the $(D-2)$-dimensional Ricci scalar intrinsic to the bifurcation surface \cite{Jacobson:1993xs}. A long calculation then gives the entropy density of the LCH in EGB theory as
\begin{align}
\rho = \rho_{B} &- 4 \alpha (\lambda_f - \lambda) \left[\frac{D-3}{D-2}(\mathcal{D}_a \mathcal{D}^a \theta) - (\mathcal{D}_a \mathcal{D}^b\sigma^a_b)\right]_{\lambda_f} \nonumber \\
&+\alpha(\lambda_f-\lambda)^2 \Bigg[ \left(\hat{R} R_{mn} - 2 \hat{R}^{ab}R_{manb}\right)k^m k^n \nonumber \\
&+2 \frac{d}{d\lambda}\left[ \frac{D-3}{D-2}(\mathcal{D}_a \mathcal{D}^a \theta) - (\mathcal{D}_a \mathcal{D}^b \sigma^a_b)\right] \nonumber \\
&+ \frac{1}{16} \delta^{a_1 a_2 a_3}_{b_1 b_2 b_3} (\mathcal{D}_{a_1} B^{b_2}_{a_2})(\mathcal{D}^{b_1} B^{b_3}_{a_3})\Bigg]_{\lambda_f}, 
\end{align}
where
$\mathcal{D}$ is the covariant derivative of the metric $\gamma_{ab}$ which is intrinsic to the $(D-2)$-dimensional constant $\lambda$ slice of the LCH, $\hat{R}$ and $\hat{R}_{ab}$ are the intrinsic Ricci curvatures of this slice, $B_{ab} = \dfrac{\theta \gamma_{ab}}{(D-2)}+\sigma_{ab}$ and $\delta$ is the Kronecker delta which is antisymmetric in all its indices. One might think that the spatial gradients of $\theta$ and $\sigma_{ab}$ can  be set to  zero on the $\lambda_f$ slice of LCH if we have a small enough patch, but this is fallacious because no matter how small the width of the patch is, the spatial gradients of $\theta$ and $\sigma_{ab}$ are related to the curvature components and there is no reason why we should be imposing any condition on the geometry.
\section{Discussion}
\label{discussion}
 Few comments on our proposal for the entropy density associated with the LCHs are in order now. First of all, for general relativity our entropy formula gives the usual area proportionality. Also, on the final equilibrium slice ($\lambda_f$) our entropy density $\rho$ agrees with $\rho_{B}$, the entropy density of the stationary black-hole horizons in the theory. 
In fact, this is true even for an arbitrary $\lambda$, provided that the expression represented by $\rho_{B}$ satisfies the physical-process version of the first law of black-hole thermodynamics. In order to see this, we note that all the derivatives of $\rho_B$ with respect to $\lambda$ vanish on any stationary slice of the black-hole horizon. Furthermore, by stationarity  $\theta$,  $\sigma$ and $d\theta/d\lambda$ on a stationary slice are all zero too, so $R_{ab} k^a k^b$ is zero by the Raychaudhari equation. Finally, if $\rho_B$ satisfies a physical process law, there can not be any flux of matter stress tensor across the stationary slice and hence $T_{ab}k^a k^b$ is zero which implies, by the equation of motion, that $E_{ab}k^a k^b $ is zero. Therefore, we see from Eq.~\eqref{eq:rho} that for any stationary slice of the black-hole horizon, we have $\rho = \rho_B$. This immediately shows that even for an arbitrary cross-section of a non-stationary black hole, our entropy formula gives $\rho = \rho_B$ as long as the perturbed black hole finally settles down to a stationary configuration in the asymptotic future.\par
What about the dependence of our entropy density on the affine-parameter $\lambda$ and on the choice of the null vector $k^a$? Recall that in hydrodynamics there is no derivation of the expression for a non-equilibrium entropy. One introduces the viscosity coefficients by hand to make sure that the entropy production is positive in the system's approach to the equilibrium. The same is true here. We have introduced terms by hand which ensure that the Clausius relation is satisfied and these terms are independent of the additive and the multiplicative ambiguity in the choice of the affine parameter $\lambda$. Furthermore, if the matter satisfies the null-energy condition then the validity of the Clausius relation immediately implies the validity of the second law of thermodynamics for quasi-stationary approach to equilibrium. Therefore, in our case entropy production is guaranteed to be positive if the matter satisfies the null-energy condition. It would be nice to understand if the $\lambda$-dependent terms in our entropy density have an interpretation in terms of the transport coefficients in hydrodynamics. \par
We should compare our entropy density to that proposed in Ref.~\cite{Guedens:2011dy}. The situation in our proposal is better than the ``Noetheresque" proposal of Ref.~\cite{Guedens:2011dy}. There the entropy density depended upon the choice of the approximate boost Killing vector field $\xi^a$ and one had to make sure that the Killing identity is satisfied to an appropriate order for the derivation to go through. In contrast, our entropy density depends upon the null generator $k^a$ of the LCH (relevant portion of) $\Sigma$ which we chose to begin with such that its expansion and shear vanish at the equilibrium point $p$ and we do not have to worry about the Killing identity at all. In both the cases though, the Clausius relation applies only to the complete patch of LCH between $\lambda = 0$ and $\lambda_f$. This is precisely analogous to what happens in the physical process version of the first-law for black holes. \par
How sensitive is our derivation of the Clausius relation to the choice of the equilibrium condition $\theta(\lambda_f)= 0$,  $\sigma^{ab}(\lambda_f) = 0$? While $\theta(\lambda_f) = 0$ is essential for our derivation, we can easily relax the condition $\sigma(\lambda_f)=0$. This results in an extra term in Clausius relation that has a natural interpretation as an internal entropy production term $\sim \sigma^{ab} \sigma_{ab}$ whose coefficient can be interpreted as the shear viscosity \cite{Eling:2006aw}. For black holes in general relativity and f(R) gravity these are precisely the terms which account for the tidal heating \cite{Chirco:2010sw}. To the best of our knowledge the expression for the shear viscosity in general theories of gravity is not known. \par 
Finally, we should mention that a physical interpretation of $\rho$ has not emerged from our construction. We have a quantity that satisfies the Clausius relation for local causal horizons in all diffeomorphism invariant theories of gravity. But is there a derivation of this expression from some basic principles?  As far as this question is concerned, the situation is better in the proposal of Ref.~\cite{Guedens:2011dy} because that construction is based upon Wald's Noether charge entropy of the black holes in the theory. It remains to be seen if a similar interpretation is possible for our proposal of the entropy density of the LCHs.

\section{Acknowledgments}
We thank Stefano Liberati for stimulating discussions and detailed comments on an earlier draft of this paper. We would also like to thank Ted Jacobson for comments. AM thanks the Institute of Mathematical Sciences, India, for hosting him during the initial stages of this work. SS's research is partially supported by the IIT Gandhinagar internal
project grant: IP/IITGN/PHY/SS/2013-001.


\begin{thebibliography}{100}  
 \bibitem{Bardeen:1973gs}
  J.~M.~Bardeen, B.~Carter, S.~W.~Hawking,
  ``The Four laws of black hole mechanics,''
  Commun.\ Math.\ Phys.\  {\bf 31}, 161-170 (1973).

 
\bibitem{Wald:1993nt}
  R.~M.~Wald,
  ``Black hole entropy is the Noether charge,''
  Phys.\ Rev.\  {\bf D48}, 3427-3431 (1993).
  [gr-qc/9307038].
  
\bibitem{Iyer:1994ys}
  V.~Iyer and R.~M.~Wald,
  ``Some properties of Noether charge and a proposal for dynamical black hole
  entropy,''
  Phys.\ Rev.\  D {\bf 50}, 846 (1994)
  [arXiv:gr-qc/9403028].

 
\bibitem{Chatterjee:2011wj}
  A.~Chatterjee and S.~Sarkar,
  ``Physical process first law and increase of horizon entropy for black holes in Einstein-Gauss-Bonnet gravity,''
  Phys.\ Rev.\ Lett.\  {\bf 108}, 091301 (2012)
  [arXiv:1111.3021 [gr-qc]].

\bibitem{Kolekar:2012tq}
  S.~Kolekar, T.~Padmanabhan and S.~Sarkar,
  ``Entropy increase during physical processes for black holes in Lanczos-Lovelock gravity,''
  arXiv:1201.2947 [gr-qc].
  
\bibitem{Jacobson:1995ab}
  T.~Jacobson,
  ``Thermodynamics of space-time: The Einstein equation of state,''
  Phys.\ Rev.\ Lett.\  {\bf 75}, 1260 (1995)
  [arXiv:gr-qc/9504004].
  
\bibitem{Padmanabhan:2009vy}
  T.~Padmanabhan,
  ``Thermodynamical Aspects of Gravity: New insights,''
  Rept.\ Prog.\ Phys.\  {\bf 73}, 046901 (2010).
  [arXiv:0911.5004 [gr-qc]].
  
\bibitem{Eling:2006aw}
  C.~Eling, R.~Guedens and T.~Jacobson,
  ``Non-equilibrium Thermodynamics of Spacetime,''
  Phys.\ Rev.\ Lett.\  {\bf 96}, 121301 (2006)
  [arXiv:gr-qc/0602001].
 
  
\bibitem{Chirco:2010sw}
  G.~Chirco, C.~Eling and S.~Liberati,
  ``Reversible and Irreversible Spacetime Thermodynamics for General
  Brans-Dicke Theories,''
  Phys.\ Rev.\  D {\bf 83}, 024032 (2011)
  [arXiv:1011.1405 [gr-qc]].
  
\bibitem{Elizalde:2008pv}
  E.~Elizalde and P.~J.~Silva,
  ``F(R) gravity equation of state,''
  Phys.\ Rev.\  D {\bf 78}, 061501 (2008)
  [arXiv:0804.3721 [hep-th]].
  
\bibitem{Brustein:2009hy}
  R.~Brustein and M.~Hadad,
  ``The Einstein equations for generalized theories of gravity and the
  thermodynamic relation $\delta Q = T \delta S$ are equivalent,''
  Phys.\ Rev.\ Lett.\  {\bf 103}, 101301 (2009)
  [arXiv:0903.0823 [hep-th]].
  
\bibitem{Parikh:2009qs}
  M.~K.~Parikh and S.~Sarkar,
  ``Beyond the Einstein Equation of State: Wald Entropy and Thermodynamical
  Gravity,''
  arXiv:0903.1176 [hep-th].
  
\bibitem{Guedens:2011dy} 
  R.~Guedens, T.~Jacobson and S.~Sarkar,
  ``Horizon entropy and higher curvature equations of state,''
  Phys.\ Rev.\ D {\bf 85}, 064017 (2012)
  [arXiv:1112.6215 [gr-qc]].
  
\bibitem{Guedens:2012sz} 
  R.~Guedens,
  ``Locally inertial null normal coordinates,''
  arXiv:1201.0542 [gr-qc].
  
 
\bibitem{Jacobson:2003wv}
  T.~Jacobson and R.~Parentani,
  ``Horizon entropy,''
  Found.\ Phys.\  {\bf 33}, 323 (2003)
  [arXiv:gr-qc/0302099].

\bibitem{Jacobson:1993xs} 
  T.~Jacobson and R.~C.~Myers,
  ``Black hole entropy and higher curvature interactions,''
  Phys.\ Rev.\ Lett.\  {\bf 70}, 3684 (1993)
  [hep-th/9305016].

\bibitem{Amsel:2007mh}
  A.~J.~Amsel, D.~Marolf and A.~Virmani,
  ``The Physical Process First Law for Bifurcate Killing Horizons,''
  Phys.\ Rev.\  D {\bf 77}, 024011 

\bibitem{Wald:1984rg}
  R.~M.~Wald,
  ``General Relativity,''
{\it  Chicago, Usa: Univ. Pr. ( 1984) 491p}

\end{thebibliography}
\end{document}